**The reliability of the gender Implicit Association Test (gIAT) for high-ability careers**

S. Stanley Young[1] and Warren B. Kindzierski[2]

[1] CGStat, Raleigh, NC, USA
[2] Independent consultant, St Albert, Alberta, Canada
Correspondence: Warren B. Kindzierski, Email: warrenk@ualberta.ca

**Abstract**
Males outnumber females in many high-ability careers in the fields of science, technology, engineering, and mathematics, STEM, and academic medicine, to name a few. These differences are often attributed to subconscious bias as measured by the gender Implicit Association Test, gIAT. We compute p-value plots for results from two meta-analyses, one examines the predictive power of gIAT, and the other examines the predictive power of vocational interests, i.e. personal interests, and behaviors, for explaining gender differences in high-ability careers. The results are clear, the gender Implicit Association Test provides little or no information on male versus female differences, whereas vocational interests are strongly predictive. Researchers of implicit bias should expand their modeling to include additional relevant covariates. In short, these meta-analyses provide no support for the gender Implicit Association Test influencing choice and gender differences of high-ability careers.

**Summary**
Males outnumber females in many high-ability careers, e.g., in the fields of science, technology, engineering, and mathematics (STEM) and academic medicine. These differences are often attributed to implicit (subconscious) bias. One objective of this study was to use statistical p-value plots to independently test the ability to reproduce the research claim of implicit bias (group tendency to prefer males over females) made in a meta-analysis of gender bias studies. The meta-analysis examined correlations between implicit bias measures based on the gender Implicit Association Test (gIAT) and measures of female and male behavior.

Real gender differences in STEM interests were noted in a meta-analysis of people's vocational interests (personal interests and behaviors that influence career choice). A second objective was to use a p-value plot to independently test the ability to reproduce sex (female−male) differences in vocational interests reported in the meta-analysis.

The p-value plots for the gIAT meta-analysis did not reproduce the research claim of implicit bias (group tendency to prefer males over females). These findings reinforce the lack of correlation between gIAT (implicit bias) measures and real-world gender behaviors in high ability careers. The p-value plot for the meta-analysis of vocational interests, in effect, supported real, non-random sex (female−male) differences in vocational interests.

There are no easy answers or fixes for gender differences in professional careers. Cultural, social, and biological factors that weigh in favor of males make it predictable that there will be more males in high-ability careers. Implicit bias measures have little or no explanatory power for gender differences in high-ability careers. There is little need to appeal to implicit bias (gIAT) measures to explain fewer females in these positions.

## Introduction

### Background

Males outnumber females in many high-status, high-tech fields (high-ability careers), e.g., science, technology, engineering, and mathematics (STEM) (Stewart-Williams & Halsey 2021) and professors in academic medicine (Van den Brink 2011). It is often assumed that females and males are nearly equal or equal in all relevant aspects of ability and interest (e.g., Hyde 2005, Hyde et al. 2009, 2019, Bosak & Kulich 2023). So, gender differences in STEM and medical professorship careers have been attributed to implicit (unconscious or subconscious) bias (Girod et al. 2016, Farrell & McHugh 2017, Hui et al. 2020).

The Implicit Association Test (IAT) is the tool developed by psychologists to measure implicit bias (Greenwald et al. 1998). It is a visual and speed reaction test done with a computer in which a person associates words with pictures. The IAT claims to measure implicit bias towards a topic of interest, such as gender bias – for example, a group tendency to prefer males over females. Gender implicit bias is measured using the gender Implicit Association Test, gIAT (Greenwald & Farnham 2000, Aidman & Carroll 2003). Since gIAT scores indicating a gender difference gap are reported to be large, it has been assumed that implicit bias is an important factor contributing to this gap (Zitelny et al. 2017).

There are three aspects that need to be examined to evaluate the accuracy of this current paradigm. First, how repeatable is the gIAT? Second, does the gIAT correlate well with explicit measures of gender difference and measurements/observations of real-world gender actions? Third, how much gender difference variance is accounted for by the gIAT? As to why these issues merit further attention, there is considerable evidence that the IAT measures in general and gIAT measures, in particular, poorly correlate with explicit measures and explain very little of the variance of gender differences (Young & Kindzierski 2024a).

### Study Objectives

Kurdi et al. (2019) and Kurdi & Banaji (2019) recently undertook a meta-analysis of studies examining correlations between implicit social cognition (i.e., implicit bias measures based on the IAT, including implicit gender bias) and criterion (real-world) measures of female and male behavior. These criterion measures included policy preference expressions, resource allocation, academic performance, subtle nonverbal behaviors, performance on interference tasks like the Stroop task, and criminal sentencing decisions. Kurdi et al. (2019) claimed that they:

> "…*found significant implicit–criterion correlations (ICCs) and explicit– criterion correlations (ECCs), with unique contributions of implicit and explicit measures revealed by structural equation modeling.*"

One objective of our study was to independently test the reliability of (ability to reproduce) a claim of implicit gender bias differences in high-ability careers. We used the publicly available Kurdi et al. and Kurdi & Banaji gender data set and statistical p-value plots (Schweder & Spjøtvoll 1982) to visually inspect the reproducibility of the claim. This objective is reasonable as Kurdi et al. (2019) themselves questioned the reliability of the IAT, including the gIAT:

> "*Statistically, the high degree of heterogeneity suggests that any single point estimate of the implicit–criterion relationship* [ICC] *would be misleading. Conceptually, it suggests* *that debates about whether implicit cognition and behavior are related to each other are unlikely to offer any meaningful conclusions.*"

If the gIAT poorly predicts gender differences and high IAT heterogeneity renders individual predictions questionable, it poses a question of what factors or variables may be important and whether IAT users have accounted for these factors in their analysis and modeling. One such set of factors may be vocational interests. Vocational interests have been defined as (Rounds & Su, 2014): *trait-like preferences to engage in activities, contexts in which activities occur, or outcomes associated with preferred activities that motivate goal-oriented behaviors and orient individuals toward certain environments.*

Su et al. (2009) noted real gender differences in STEM interests in a meta-analysis of vocational interests; with the most important interest identified as… *interests in working with things versus working with people* (referred to as 'things−people interests'). They observed that these differences paralleled the female-male composition in STEM educational programs and occupations and may play a role in gender occupational choices and gender disparity in the STEM fields. This also may be the case for other high-ability careers like academic medicine.

A second objective of our study was to independently test the ability to reproduce sex (female−male) differences in vocational interests reported in the Su et al. (2009) meta-analysis. Specifically, we used the Su et al (2009) data set and statistical a p-value plot to visually inspect the reproducibility of their claim.

**Methods**
*gIAT*
We developed and posted a research plan for our study at Researchers.One (Young & Kindzierski 2024b). The three instruments (variables) of interest regarding the validity of an IAT measurement of gender bias include: the gIAT (A); explicit measure(s) of bias, e.g., indications of attitude, belief, or preference of gender bias captured in a questionnaire or similar research instrument (B); and criterion measures – observations/measurements of real-world gender biased behaviors or actions (C). Measures of 'A' and 'B' should be positively correlated with 'C' for 'A' and 'B' to be considered valid.

Meta-analysis is used across multiple fields, including psychology. It is a procedure for combining test statistics from individual studies that examine a particular research question (Egger et al. 2001). A meta-analysis can evaluate a research question by taking a test statistic (e.g., a correlation coefficient between two variables of interest) along with a measure of its reliability (e.g., confidence interval) from multiple individual studies from the literature. The test statistics are combined to give, theoretically, a more reliable estimate of correlation between the two variables.

A requirement of meta-analysis is that test statistics taken from individual studies for analysis are unbiased estimates (Boos & Stefanski 2013). Given this requirement/assertion, independent evaluation of published meta-analysis on a particular research question can be used to assess the statistical reproducibility of a claim coming from that field of research (Young & Kindzierski 2019, Kindzierski et al. 2021, Young & Kindzierski 2022, Young & Kindzierski 2023).

We examined and extracted meta-analysis data sets from Kurdi et al. (2019) and Kurdi & Banaji (2019) meta-analysis. Studies initially selected were those dealing with gender. These studies selected had a combined sample size of 1,155.

The gender data of interest to us consisted of individual study author, year, title, journal, and correlation coefficient (*r*) values for three two-variable comparisons:

- implicit–criterion correlations (ICCs) – an ICC is the correlation between a gIAT measurement/score and observations/measurements of real-world gender-related measures, i.e., gender attitude, stereotype, and identity.
- explicit–criterion correlations (ECCs) – an ECC is the correlation between an explicit measure(s) of bias, e.g., indications of attitude, belief, or preference of gender bias captured in a questionnaire or similar research instrument and real-world gender-related measures.
- implicit–explicit correlations (IECs) – an IEC is the correlation between implicit and explicit variable measurements.

Only studies having data for all ICC, ECC, IEC comparisons were used for our evaluation. This smaller data set comprised a sample size of 535 from 27 individual studies. Mean ICC, ECC, and IEC correlation coefficient (*r*) values were computed for each of the 27 studies.

We then converted *r* values to p-values using Fisher's *Z*-transformation (Fisher 1921, Wicklin 2017). Fisher's *Z*-transformation is a statistical technique used to compute a p-value for a correlation coefficient *r* given sample size *n*. The formula for Fisher's *Z*-transformation is:

$$Z = arctanh\ (r) = 0.5 \times ln[(1+r)/(1-r)], \hspace{3cm} [1]$$

which is considered to follow a normal distribution with standard error (*SE*) of:

$$SE = SQRT[1/(n-3)]. \hspace{3cm} [2]$$

Average *Z-score*/*SE* values were converted back to p-values using the standard normal distribution (Fisher et al. 1990).

*Vocational Interests*

Su et al. (2009) undertook a meta-analysis examining the magnitude and variability of sex differences in vocational interests. Technical manuals of vocational interest inventories from educational settings, public institutions, and private organizations (n=108) were initially screened and evaluated, of which 47 interest inventories were selected for meta-analysis. The inventories were intended to measure vocational interests (documented assessments of people's interests in potential careers, educational paths, and the world of work). The inventories were published in English with collections of scores from female and male populations (norm samples) from the US or combined norm samples from both the US and Canada.

The 47 inventories were published over four decades (between 1964 and 2007) and comprised a total of 81 samples consisting of 259,518 women and 243,670 men ($n_{total}$ = 503,188). Mean ages of the samples ranged from 12.5 to 42.6 years. The oldest cohort of the samples was born in 1939, and the youngest in 1987.

Su et al. calculated weighted mean effect sizes (Cohen's *d*), standard deviation (SD) and lower- & upper-95 percent confidence intervals (*CI*s) for eleven different dimensions of vocational interests. These included: [1] interest in working with things versus people (Things-People), [2] interest in working with data versus ideas (Data-Ideas), RIASEC interests ([3] **R**ealistic interest in working with things and gadgets or working outdoors, [4] **I**nvestigative interest in science, including mathematics, physical and social sciences, and biological and medical sciences, [5] **A**rtistic interest in creative expression, including writing and the visual and performing arts, [6] **S**ocial interest in helping people, [7] **E**nterprising interest in working in leadership or persuasive roles directed toward achieving economic objectives, [8] **C**onventional interest in working in well-structured environments, especially business settings, and STEM interests ([9] Science, [10] Mathematics, [11] Engineering).

We used *d* values and lower- & upper-95 percent *CI*s to estimate standard error (*SE*) and *Z*-scores for each dimension assuming normal distributions: $SE = (95\% \, CI - 5\% \, CI)/3.92$ and $Z = d/SE$. *Z-score* values were then converted to p-values using the standard normal distribution (Fisher et al. 1990).

*p−value Plot*

The p-values for a set of test statistics were displayed in a p-value plot. The plot is used to visually check characteristics of test statistics addressing the same research question. The plot – originally presented by Schweder & Spjøtvoll (1982) – is well-regarded, being cited more than 500 times in scientific literature (Google Scholar 2024).

In our case, the plot is used to assess whether the p-values follow a uniform distribution… if there is little or no correlation (i.e., nothing is going on) between two variables, the p-values for a data set should be evenly spread over the interval 0 to 1. Specifically, a p-value plot is a 2-way scatterplot where the observed p-values are ordered from smallest to largest and plotted against the integers: 1, 2, 3…. *n*. Under the null hypothesis (i.e., nothing is going on between two variables), the uniform distribution presents as a near−45º line. A layperson's explanation of this is provided in Appendix 2 of Young et al. (2021).

There are several ways to interpret p-value plots (Young & Kindzierski 2019, Kindzierski et al. 2021, Young et al. 2021, Young & Kindzierski 2022, Young & Kindzierski 2023):

- If the observed data points (p-values) fall approximately on a near−45º line in the plot, it suggests a good fit with the theoretical (uniform) distribution and a null (no) association between tested variables.
- If the observed data points are mostly less than 0.05 and fall on a line with a shallow slope in the plot, there could be a real, non-random association between tested variables.

Regarding gIAT p-value plots, another possibility is that in the absence of methodological, reporting, and publication biases (e.g., Ioannidis 2005, 2022, Schimmack 2021), deviations from a near−45º line in the plot may indicate departures from the uniform distribution and that there could be a real, non-random association between tested variables. This possibility is considered remote as the methodological limitations (and biases) of the IAT are well-published (e.g., Arkes & Tetlock 2004, Blanton & Jaccard 2006, Tetlock & Mitchell 2009, Oswald et al. 2013, 2015, Mitchell & Tetlock 2017, Schimmack 2019 and 2021).

*Volcano Plot*

A volcano plot was used to visually examine p-values from the Su et al. (2009) meta-analysis data set. This is a type of scatterplot used to check for patterns in data, particularly for identifying statistically significant differences between two populations (Li et al., 2014; Hur et al., 2018). The volcano plot we used was constructed by graphing the negative log10 of the p-value on the y axis against the *d* value on the x axis. A consistent data set in a volcano plot should resemble an erupting volcano. Data points with low p-values (highly statistically significant results) appear toward the top of the plot. Data points in the top-right and top-left area of a volcano plot are of interest because they are the most different between the two conditions of interest – i.e., results with *d*s > 0 on the right versus *d*s < 0 on the left.

## Results

*gIAT*

We examined three correlations of interest with the Kurdi et al. (2019) and Kurdi & Banaji (2019) meta-analysis gender data set – i) implicit–criterion correlations (ICCs), ii) explicit–criterion correlations (ECCs), and implicit–explicit correlations (IECs). Figure 1 displays *Z* statistic frequency histograms (left side), box and whisker plots (right side), and quantiles (bottom) for each of the three correlations. Medians for each of the three comparisons are remarkably close to zero, which would make sense if the IAT is invalid measure.

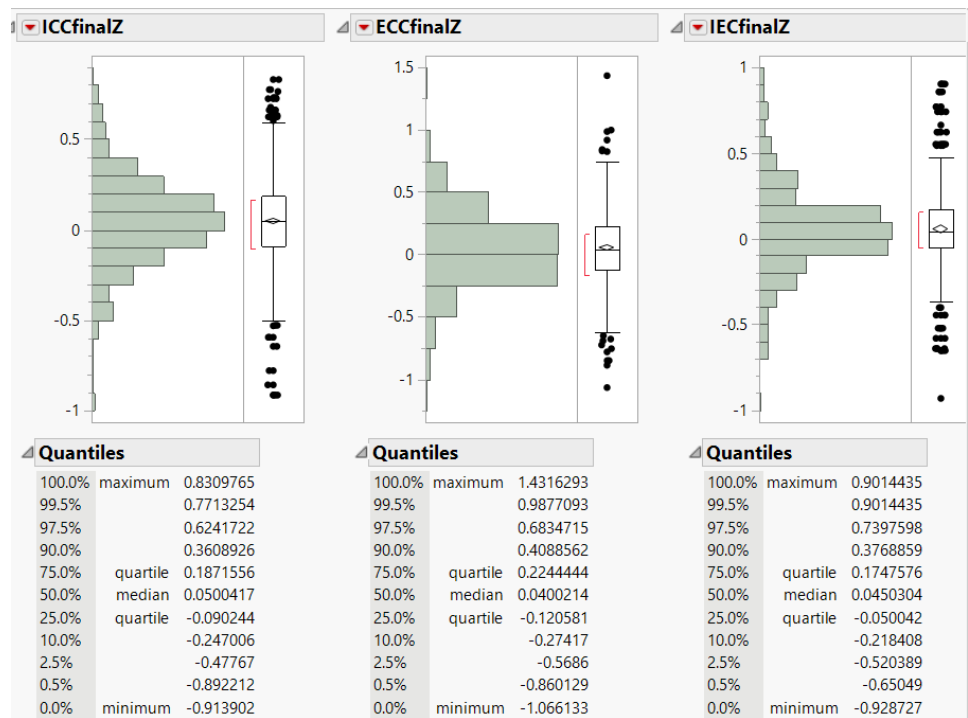

*Figure 1*. Frequency histograms (left side), box and whisker plots (right side), and quantiles (bottom) for Z-statistics from three correlations – ICC, ECC, & IEC – for the gender data sets of Kurdi et al. (2019) and Kurdi & Banaji (2019). Note: ICCfinalZ=correlation between implicit & criterion measures, ECCfinalZ=correlation between explicit & criterion measures, IECfinalZ=correlation between implicit & explicit measures.

Rank-ordered p-values computed for 27 ICC, ECC, and IEC correlations are presented in Figures 2 to 4, respectively for 27 studies from Kurdi et al. (2019) and Kurdi & Banaji (2019) meta-analysis dealing with gender. The p-value trends displayed irregular (unexpected) shapes in each of the plots. Averaging $r$ values in each individual study reduces the measurement error, which may explain the resulting irregular trend of p-values in these plots. In any case, p-values in the plots are all greater than 0.05 and do not support real associations between the tested variables.

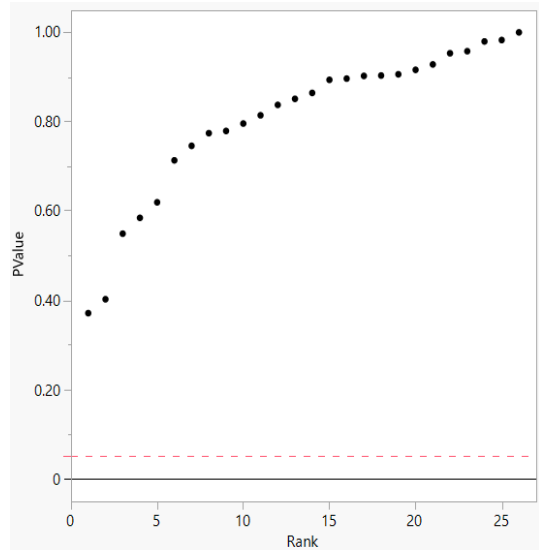

*Figure 2*. Rank-ordered p-values computed for 27 ICC (implicit−criteria measure) correlations from the Kurdi et al. (2019) and Kurdi & Banaji (2019) meta-analysis dealing with gender. Note: p-values were computed from mean correlation coefficient ($r$) values for each study.

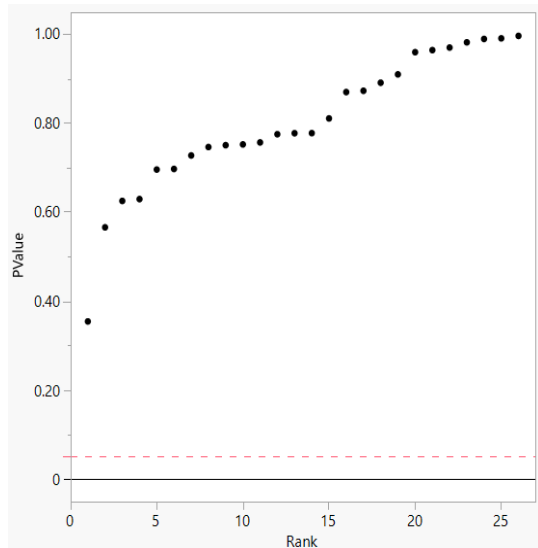

*Figure 3*. Rank-ordered p-values computed for 27 ECC (explicit−criteria measure) correlations from the Kurdi et al. (2019) and Kurdi & Banaji (2019) meta-analysis dealing with gender. Note: p-values were computed from mean correlation coefficient ($r$) values for each study.

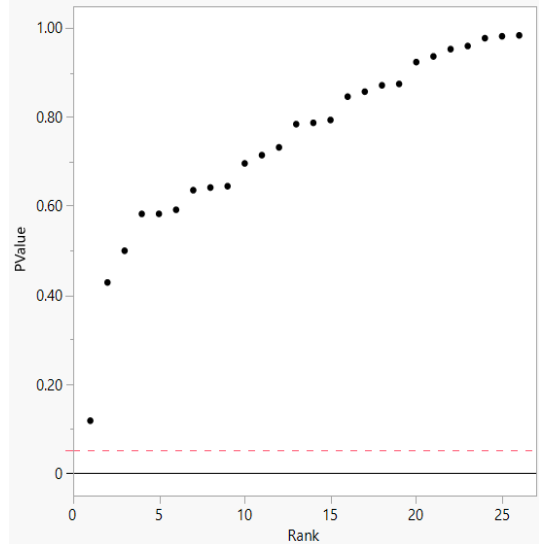

*Figure 4.* Rank-ordered p-values computed for 27 IEC (implicit−explicit measure) correlations from the Kurdi et al. (2019) and Kurdi & Banaji (2019) meta-analysis dealing with gender. Note: p-values were computed from mean correlation coefficient (*r*) values for each study

*Vocational Interests*

Rank-ordered p-values computed for the 11 vocational interest dimensions reported by Su et al. (2009) are shown in Figure 5a. The plot shows that most of the observed data points are less than 0.05. The plot, in effect, supports real, non-random sex (female−male) differences in vocational interests. This is more apparent by examining the volcano plot (Figure 5b). Figure 5b shows Social, Artistic, Conventional, and Data-Ideas vocational interest dimensions favoring women, with the Social and Artistic dimensions being the strongest. While the strongest vocational interest dimensions favoring men are Realistic, Things-People, and Engineering.

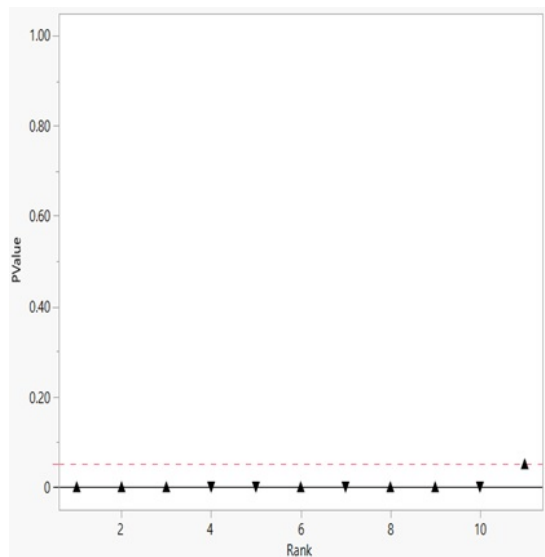

*Figure 5a.* Rank-ordered p-values computed for 11 different vocational interest dimensions reported by Su et al. (2009). Note: (▼) vocational interest dimension favoring females; (▲) vocational interest dimension favoring males.

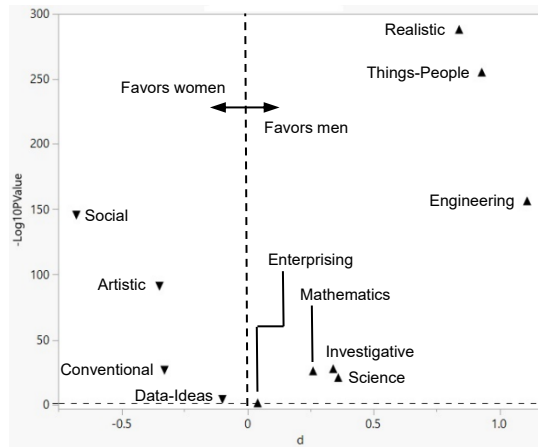

*Figure 5b.* Volcano plot for 11 different vocational interest dimensions reported by Su et al. (2009). Note: (▼) vocational interest dimension favoring females; (▲) vocational interest dimension favoring males.

## Discussion

We considered three instruments of interest to the validity of a gIAT measure – the gIAT (implicit measures); explicit measure(s) of bias, e.g., indications of attitude, belief, or preference of gender bias captured in a questionnaire or similar research instrument; and observations/measurements of real-world gender biased behaviors or actions (criterion measures). These instruments were considered in relation to their applicability to explain gender differences in high-ability careers. The p-value plots (Figures 2 to 4) explored implicit–criterion correlations (ICCs), explicit–criterion correlations (ECCs), and implicit–explicit correlations (IECs). The data points (p-values) in these plots were all greater than 0.05 and do not support real associations between the tested variables.

Implicit bias, as measured by the Implicit Association Test, IAT, is one of psychology's biggest ideas in the last 30 years. A Google Scholar search of the phrase "Implicit Association Test" on 2 May 2024 returned 49,500 articles and/or citations. Even so, there is now strong criticism of the IAT. Two key criticisms predominate – validity and reliability. Regarding validity, research has shown that the IAT has low correlation with explicit measures and real-world actions (Fiedler et al. 2006, Blanton et al. 2009, Oswald et al. 2013 and 2015, Schimmack 2019 and 2021, Mitchell & Tetlock 2020). This research implies that the IAT does not measure what it is said to measure.

Reliability is also in question. Repeated measures of the IAT shows that much of the 'measurement' is measurement error; the value for an individual fluctuates considerably around a mean value (Schimmock 2021), so much so that is not useful for, in the case of race, predicting discrimination. The p-value plots (Figures 2 to 4) support these IAT criticisms. The applicability of the gIAT for describing gender differences in high-ability careers is questionable.

We previously stated that there is evidence that the IAT in general and the gIAT, in particular, poorly correlate with explicit measures and explain very little of the variance of gender differences (Young & Kindzierski 2024a). This begs the question of what is missing in studies of gender differences in high-ability careers? As noted previously, Su et al. (2009) observed real gender differences in STEM interests in a meta-analysis of vocational interests. The p-value and volcano plots (Figures 5a and 5b) show strong female−male differences for several vocational interest dimensions (Social and Artistic dimensions favoring women; and Realistic, Things-People, and Engineering dimensions favoring men).

Given these strong female−male differences, there is merit in examining aspects of bias in the gIAT – specifically, that of residual bias from possible missing confounding factors. Good examples being female−male differences in vocational interests. This bias is referred to as *omitted variable bias* in literature (Wilms et al 2021, Hirukawa et al. 2023, Basu 2024). A complete discussion of all possible factors in relation to STEM and medical academic careers is not something that can be addressed here. However, examples for several factors are used to show the importance bias from omitted variables.

*Confounders*
Initially we use simple linear regression models between two variables to examine theoretical aspects of confounding from omitted variables. We then follow up with several gender studies of professors in academic medicine that show the importance of confounding from omitted variables. Linear regression is widely used to help understand real-world situations, for example, the association between one predictor variable and one outcome (or response) variable. Keep in mind that a regression model is just that, a model. Once the model is made, it is still a model, an approximation to the world.

First, let's consider simple linear regression models; one model for females and another for males:

Female:  $Y_f = \beta_0 + \beta_1 X_{1f} + \varepsilon$ [3]
Male:  $Y_m = \beta_0 + \beta_1 X_{1m} + \varepsilon$ [4]

The above models say that an outcome, Y, is predictable using a linear additive function of an intercept, $\beta_0$, and $\beta_1 X_1$, and there is a random error in the prediction captured by $\varepsilon$. Y is an outcome variable and X is a predictor variable of interest. The subscripts "f" and "m" denote females and males, respectively.

Now let's consider a specific example of the percentage of full professors in medical schools. Let Y be the percentage of full professors in medical schools; let X be a predictor variable. The subscripts "f" and "m" denote females and males, respectively. The regression models can be expanded by adding more variable (X) terms:

Female:  $Y_f = \beta_0 + \beta_1 X_{1f} + \beta_2 X_{2f} + \beta_3 X_{3f} + \beta_4 X_{4f} + \ldots + \beta_p X_{pf} + \varepsilon$ [5]
Male:  $Y_m = \beta_0 + \beta_1 X_{1m} + \beta_2 X_{2m} + \beta_3 X_{3m} + \beta_4 X_{4m} + \ldots + \beta_p X_{pm} + \varepsilon$ [6]

The above mathematical models indicate that percentage of female/male full professors in medical schools, Y, is predictable using a linear additive function of the variables $X_1$, $X_2$, $X_3$, $X_4$, $\ldots X_p$ specific to females (or males) and a random error term $\varepsilon$. Here, $X_1$ is the predictor variable of interest and the other variables – $X_2$, $X_3$, $X_4$, $\ldots X_p$ – are said to be confounding factors. As before the subscripts "f" and "m" denote females and males, respectively.

Let's say we are interested in looking at differences between female and male full professors in medical schools, $Y_f − Y_m$ (i.e., Eqn. [5] − Eqn. [6]). Using a difference between two multiple linear regression models, Young (2008) showed that there is the potential for residual bias in multiple linear regression if the models omit important terms (i.e., unknown confounders):

$$(Y_f - Y_m) - [\text{known confounders, i.e., } X_2, X_3, X_4, ...X_p] = \beta_1(X_{1f} - X_{1m}) + [\text{unknown confounders}] \qquad [7]$$

If there are important unknown confounders omitted in modeling, then a modeling exercise exploring gender differences can be biased (unreliable). Two gender studies of professors in medical schools are considered below – Jena et al. (2015) and Carr et al. (2018).

Jena et al. (2015) – Jena et al. examined the proportion of females at the rank of full professor in US medical schools over the period 1980−2014. Their sample size was very large – 91,073 US academic physicians. They considered physician gender (independent or predictor variable) and several confounding factors in their model – age, years since residency, specialty, authored publications (a measure of research productivity), National Institute of Health (NIH) funding, and clinical trial participation. They noted that the percentage of full professors was 28.6 (males) and 11.9 (females) before adjusting for these variables – giving a gap of 16.6 percentage points. After adjusting for the multiple omitted variables noted above, the gap shrank to only 3.8 percentage points.

Carr et al. (2018) – Carr et al. tracked 1,273 faculty at 24 medical schools in the US for 17 years to identify predictors of advancement, retention, and leadership for female faculty as part of the National Faculty Survey. This was a national cohort of faculty followed from 1995 to 2012–2013 to examine differences in career outcomes by gender. Carr et al. (2018) looked at physician sex (independent variable) and numerous confounders in their model. These included: race, medical specialization, seniority level, percent effort distribution for administrative, research, clinical, and teaching activities, marital status, parental status, and academic productivity measured by the total number of refereed publications.

After adjusting for all confounders except refereed career publications, females were less likely than males to achieve the rank of professor (odds ratio, OR = 0.57; 95% confidence interval, CI, 0.43–0.78) or to remain in academic careers (OR = 0.68; 95% CI, 0.49–0.94). However, when total number of refereed publications was added to their model, Carr et al. (2018) observed that differences by gender in retention and attainment of senior rank were no longer significant.

The findings of Jena et al. (2015) and Carr et al. (2018) show the importance of confounders in studies of gender differences in academic medicine, including the gIAT. Any study could quite easily show large gender differences, for example, in US academic physician positions by omitting many of the confounders used by Jena et al. (2015) and Carr et al. (2018).

*Broad Confounders*
In considering professional career choices of females and males, it is relatively easy to tabulate numbers of females and males in a profession. Consider females in STEM academic faculty positions. As of 2019, females/males made up 34.5/65.5% of STEM faculty and 28.2/71.8% of tenured STEM faculty at academic institutions in the US (NSF NCSES 2019). Gender differences in STEM and medical professorship careers have been attributed to implicit bias (Girod et al. 2016, Farrell & McHugh 2017, Hui et al. 2020). However, applicability and reproducibility of the gIAT – Figures 2 to 4 – has been shown to be lacking (flawed) for describing gender differences in these careers. In essence, implicit bias is provides a poor explanation of these differences.

Haier (2009) noted that there are no simple answers to gender differences in professional careers, and that there are persuasive data that suggest multiple interacting factors are at play – some cultural and social, others biological. Several of these have meaningful contributions, including female−male differences in (Gottfredson 2003, Levy & Kimura 2009, Coyle 2018, Stewart-Williams & Halsey 2018):

- math, verbal, and social skills
- interests, career, and lifestyle preferences
- cognitive (thinking) ability

Some of these factors weigh in favor of females, others in favor of males. These factors are most likely known among psychologists, and indeed female−male interests are known to just about everyone. Yet it is puzzling that they are not acknowledged or used as confounders to see if implicit bias (i.e., gIAT) measures adds any predictive power to gender differences in the STEM field and academic medicine.

For illustrative purposes, we further explore two possible vocational interest dimension confounders that weigh in favor of males – Things-People and Engineering interests (refer to Figure 5b). We ignored the Realistic dimension as Su et al. (2009) noted that it accounted for most of the Things−People sex difference. Our interest here is to try to understand how these interests may confound an implicit bias claim based on the gIAT.

Things−People vocational interests – Females and males are free to choose their professional careers and how they want to make their way in the world based on their interests. Interests of people are predictive of their behaviors in particular environments (Rounds & Su 2014), for example, choice of college major, and career occupation. Regarding gender differences in working with things or people, there is a very strong effect in Figure 5b showing that males favor working with things over people.

A hypothetical example is illustrated with Things−People data from Su et al. (2009) – refer to the standard normal curves presented in Figure 6. Su et al. reported a mean effect size ($d$) for the Things−People dimension of 0.93 favoring men in their study. This is comparable to a Things−People $d = 1.01$ favoring men reported by Morris (2016) in a more recent study of vocational interests of US residents.

For our example in Figure 6 we assumed similar standard deviations for both female and male distributions for simplicity (i.e., these distributions are equally spread out). Males are used as the reference point in Figure 6, i.e., the distribution for males have mean, $\mu = 0$ and standard deviation, $\sigma = 1$; while females have $\mu = −0.93$ (after Su et al., 2009) and $\sigma = 1.00$ (assumed). The distribution for females is shifted slightly to the left in the figure. A shift to the right of zero indicates greater interest for things over people, whereas a shift to the left of zero indicates greater interest for people over things.

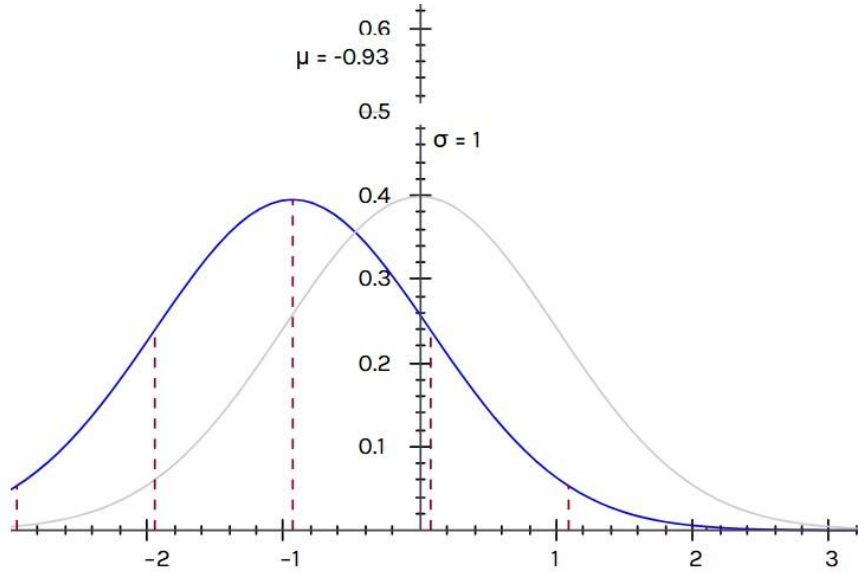

*Figure 6.* Normal, Gaussian, distribution of Things−People interests (after Su et al. 2009); males (reference): mean (μ) = 0, standard deviation (σ) = 1; females: μ = −0.93, σ = 1. Note: females (──────), males (──────).

Now, 50 percent of males are to the right of zero, whereas only 17.6 percent of females are to the right with a preference for working with things (refer to Table 1). Moving further to the right of zero on the horizontal axis in Figure 6 makes a huge difference in numbers of females with a preference for working with things. The ratio of males to females at one (two) standard deviation to the right is 5.9 (13) (Table 1).

These differences are noteworthy given that STEM fields are typically things-oriented (Su & Rounds, 2015; Stoet & Geary, 2022). As an example, if a person needs to be one to two standard deviations to the right of zero in the Things−People distribution to select a STEM vocation, then there will be 5.9 to 13 times as many males as females in this pool.

Table 1. Normal, Gaussian, distribution characteristics of Things-People interests for females and males after Su et al. (2009).

| Standard deviation (SD) relative to male | Area under male curve above SD (AUC$_M$) | Area under female curve above SD (AUC$_F$) | AUC$_M$/AUC$_F$ Ratio |
|---|---|---|---|
| 0 | 0.5000 | 0.1762 | 2.8 |
| 1 | 0.1587 | 0.0268 | 5.9 |
| 2 | 0.0228 | 0.0017 | 13 |
| 3 | 0.0014 | 0.0004 | 32 |

Engineering Vocational Interests – Students graduating High School are more likely to enroll in engineering programs if their standardized Scholastic Assessment Test (SAT) math scores are favorable (Tan et al., 2022). Further, math proficiency skills (SAT math scores) have been shown to be a strong positive predictor of attending an engineering graduate program (Ro et al., 2017); a graduate PhD degree is a necessary requirement for any type of academic position.

Math proficiency skills and cognitive ability have an overlap (Hart et al., 2009). These skills are also known to be highly general capabilities for processing complex information of any type and for successful performance in high complexity careers (Gottfredson 2002, Kuncel & Hezlett 2010). These attributes can influence job performance particularly higher up occupational hierarchies in various fields, including STEM and academic medicine.

Math proficiency skills based on scores in numeracy tests (e.g., SAT math scores) have more consistent positive effects on job content skills and wages than scores on non-math literacy tests (Hanushek et al. 2015, Kahn 2018). Female−male performance in the standardized SAT for math over a 44-year period in United States is shown in Figure 7.

Females consistently underperformed males by 30 or more points over the period in Figure 7. Further, 2016 SAT math score data (mean ($\mu$) = 494, $\sigma$ = 116, n = 875,342 (females); $\mu$ = 524, $\sigma$ =126, n = 762,247 (males)) (College Board, 2016) translates into an effect size ($d$) disadvantage of −0.248 ($\sigma$ = 0.92) for females.

How might a general female disadvantage in math proficiency skills play out in high-ability careers? A hypothetical example is illustrated with 2016 SAT math score data (refer to Figure 8). Males are again used as the reference point, i.e., the distribution for males have $\mu$ = 0 and $\sigma$ = 1, while females have $\mu$ = −0.248 (shown as −0.25) and $\sigma$ = 0.92 in Figure 8. The distribution for females is shifted slightly to the left in the figure. A shift to the right of zero indicates higher SAT math scores, whereas a shift to the left of zero indicates lower scores.

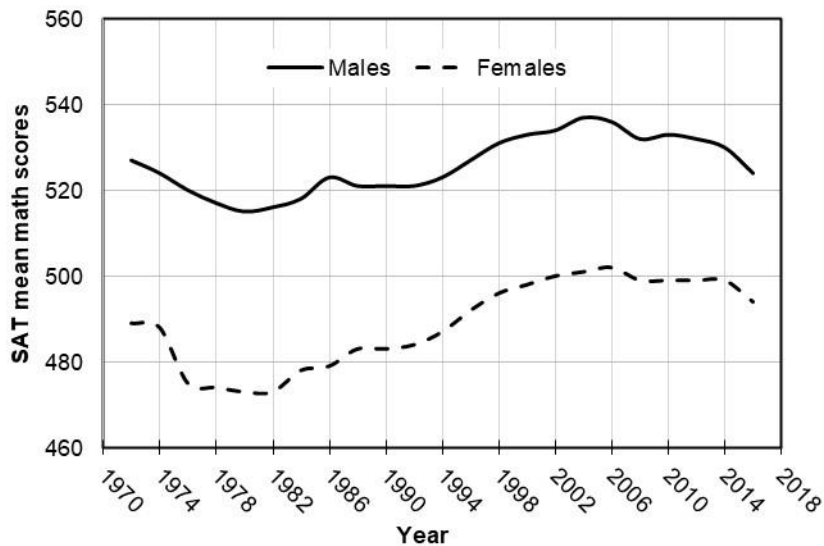

*Figure 7.* Standardized Scholastic Assessment Test (SAT) mean math scores in United States – females vs. males, 1972−2016, (College Board 2016, 2024).

Differences further above the average (i.e., above $\mu$ = 0) is where it may be relevant for females versus males competing for and occupying positions in high-ability careers. The net effect is that the ratio (and numbers) of males to females increases with increasing SAT math score – there are fewer numbers of females than males at higher levels of SAT math scores.

For the area under the curves in Figure 8 representing SAT math scores greater than one and two standard deviations to the right of zero on the horizontal axis, there are more males than females, actually 1.8 and 3 times as many (refer to Table 2). While greater than three standard deviations there are 7 times as many. At the right side of Figure 8 there are fewer females as there are males.

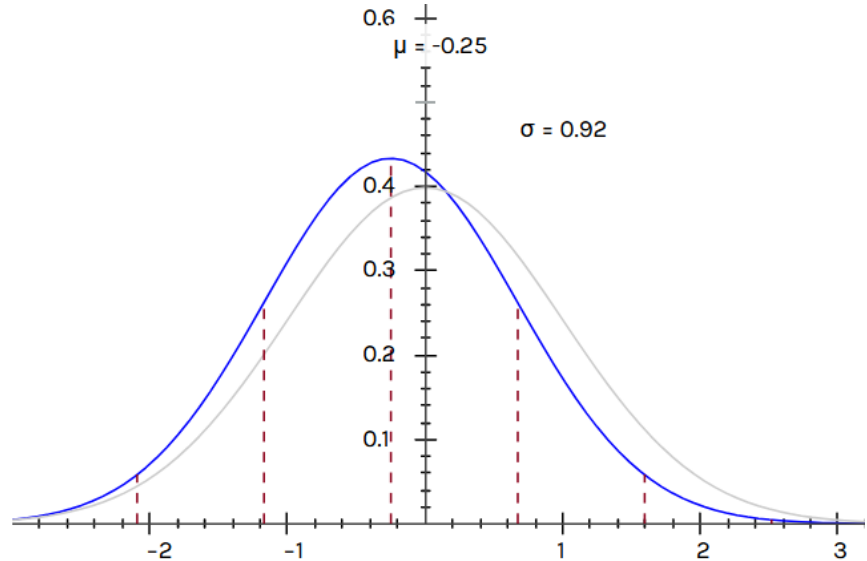

*Figure 8.* Normal, Gaussian, distribution of 2016 SAT math score data after College Board (2016); males (reference): mean (μ) = 0, standard deviation (σ) = 1; females: μ = −0.248, σ = 0.92. Note: females (———), males (———).

Table 2. Normal, Gaussian, distribution characteristics of SAT mean math scores in United States for females and males in 2016 (College Board 2016).

| Standard deviation (SD) relative to male | Area under male curve above SD ($AUC_M$) | Area under female curve above SD ($AUC_F$) | $AUC_M/AUC_F$ Ratio |
|---|---|---|---|
| 0 | 0.5000 | 0.3930 | 1.3 |
| 1 | 0.1587 | 0.0875 | 1.8 |
| 2 | 0.0228 | 0.0075 | 3.0 |
| 3 | 0.0014 | 0.0002 | 7.0 |

These two hypothetical examples illustrate that there are fewer numbers of females with Things−People interests and that math proficiency skills may offer better explanations than implicit bias for gender differences in the makeup of academic STEM or medicine faculties. Factors like this and others that weigh in favor of males make it predictable that there will be more males than females in high-ability careers.

*Equity versus Equality*
Equity normally means that everyone is treated justly and fairly, based on their demonstrated abilities and efforts as established in their history. Equality means equal. For example, the belief that females and males are equal and career opportunities for females and males need to be equal. An example of this is the European Commission 2019 Gender Equality strategy (European Commission 2021). This is a commitment to promote gender equality in research and innovation (R&I) and fuller participation of females in R&I-oriented (e.g., engineering, science) careers in the European Union.

If the makeup of female-to-male employment in R&I-oriented careers is unbalanced, does this mean that efforts need to be made so that it becomes balanced? According to the European Commission (2021)… "*There is no sustainable recovery* [in the European Union] *if it is not gender-sensitive.*" If indeed females and males are not equal in ability or employment sentiment, European Commission (2021) supports actions working towards equality of outcome. Does this mean that females with lower Things−People interests (Figure 6) or lower math proficiency skills (Figure 8) will be promoted over males of greater ability in R&I-oriented careers?

If STEM or medical faculty selection committees purposely aim for equality (equal numbers) of females and males in their faculties, there could be a selection bias toward females with lower Things−People interests and math proficiency skills. To get numbers of females equal to numbers of males in STEM and medical school faculties, selection committees would need to go deeper into the female distributions (i.e., towards the left of Figures 6 and 8). There could be a disadvantage here to females relative to males in these faculties as females progress through their careers – fewer papers, fewer promotions, etc.

Cultural, social, and biological factors that weigh in favor of males make it predictable that there will be more males than females in high-ability careers. There is little need to appeal to implicit bias to explain fewer females in these positions. There are females with interests in working with things and math proficiency skills, and other factors that weigh in their favor, and they are likely eagerly recruited; but there would not be enough to go around to create gender equality in these fields.

Virtually none of gIAT studies discuss or consider other factors that could explain fewer females in these fields. We illustrated two possible factors that favor males, interests in working with things and math proficiency skills, but there could be other important ones. Hypothetically, as one moves up the ladder at every career position requiring higher abilities, there will be fewer eligible females than males to fulfill these positions. One should not be surprised that implicit bias (gIAT) measures have little or no explanatory power for describing gender differences in high-ability careers.

**Findings**

Males outnumber females in many high-ability careers, for example, in the fields of science, technology, engineering, and mathematics (STEM) and professors in academic medicine. These differences are often attributed to implicit (subconscious) bias.

Statistical p-value plots were used to independently test the ability to reproduce a claim of implicit bias made in Kurdi et al. (2019) and Kurdi & Banaji (2019) meta-analysis of gender bias studies. The meta-analysis examined correlations between implicit bias measures based on the gender Implicit Association Test (gIAT) and measures of intergroup (female and male) behavior.

The p-value plots constructed using data sets from the meta-analysis (Figures 2 to 4) did not support real associations between the tested variables. These plots did not reproduce the research claim of implicit bias made in the Kurdi et al. (2019) and Kurdi & Banaji (2019) meta-analysis. These findings reinforce the lack of correlation between gIAT (implicit bias) measures and real-world gender behaviors in high-ability careers.

A p-value plot was used to independently test the ability to reproduce sex (female−male) differences in vocational interests reported in a meta-analysis of vocational interests by Su et al. (2009). The p-value plot for the meta-analysis (Figure 5a), in effect, supported real, non-random sex (female−male) differences in vocational interests.

There are no easy answers or fixes for gender differences in professional careers. Cultural, social, and biological factors that weigh in favor of males make it predictable that there will be more males in high-ability careers. Implicit bias measures have little or no explanatory power for gender differences in high-ability careers. There is little need to appeal to implicit bias (gIAT) measures to explain fewer females in these positions.

## References

Aidman, E. V., & Carroll, S. M. (2003). Implicit individual differences: Relationships between implicit self-esteem, gender identity, and gender attitudes. *European Journal of Personality*, 17, 19–36. https://doi.org/10.1002/per.465

American College Testing Program. (1995). *Technical manual: Revised Unisex Edition of the ACT Interest Inventory (UNIACT)*. Iowa City, IA: American College Testing Program.

Arkes, H. R., & Tetlock, P. E. (2004). Attributions of implicit prejudice, or "would Jesse Jackson 'fail' the implicit association test?" *Psychological Inquiry*, 15, 257–278. https://doi.org/10.1207/s15327965pli1504_01

Basu, D. (2024). *How likely is it that omitted variable bias will overturn your results? Working Paper No. 2024-1*. Amherst, MA: University of Massachusetts, Department of Economics. http://dx.doi.org/10.2139/ssrn.4704246

Blanton, H., & Jaccard, J. (2006). Arbitrary metrics in psychology. *American Psychologist*, 61, 27–41. https://doi.org/10.1037/0003-066X.61.1.27

Blanton, H., Jaccard, J., Klick, J., Mellers, B., Mitchell, G., & Tetlock, P. E. (2009). Strong claims and weak evidence: Reassessing the predictive validity of the IAT. *Journal of Applied Psychology*, 94(3), 567–582; Discussion 583–603. https://doi.org/10.1037/a0014665

Boos, D. D., & Stefanski, L. A. (2013). *Essential Statistical Inference: Theory and Methods*. New York, NY: Springer. https://doi.org/10.1007/978-1-4614-4818-1

Bosak, J., & Kulich, C. (2023). Gender similarities hypothesis. In: Shackelford, T.K. (ed.), *Encyclopedia of Sexual Psychology and Behavior*. Springer, Cham. https://doi.org/10.1007/978-3-031-08956-5_163-1

Carr, P. L., Raj, A., Kaplan, S. E., Terrin, N., Breeze, J. L., & Freund, K. M. (2018). Gender differences in academic medicine: Retention, rank, and leadership comparisons from the National Faculty Survey. *Academic Medicine*, 93(11), 1694–1699. https://doi.org/10.1097/ACM.0000000000002146

College Board. (2016). SAT 2016 College-Bound Seniors, Total Group Profile Report. https://reports.collegeboard.org/media/pdf/2016-total-group-sat-suite-assessments-annual-report.pdf

College Board. (2024). *SAT Suite Data and Reports Archive*. New York, NY: College Board. https://reports.collegeboard.org/sat-suite-program-results/data-archive

Coyle, T. R. (2018). Non-g factors predict educational and occupational criteria: More than g. Journal of Intelligence, 6(3), 43. https://doi.org/10.3390/jintelligence6030043

Egger, M., Davey Smith, G., & Altman, D. G. (2001). Problems and limitations in conducting systematic reviews. In: Egger, M., Davey Smith, G., & Altman, D. G. (eds.), *Systematic Reviews in Health Care: Meta‑analysis in Context, 2nd ed*. London, UK: BMJ Books. https://doi.org/10.1002/9780470693926

European Commission. (2021). *She Figures 2021, Gender in Research and Innovation Statistics and Indicators*. European Commission, Directorate-General for Research and Innovation, Brussels. https://data.europa.eu/doi/10.2777/06090

Farrell, L., & McHugh, L. (2017). Examining gender-STEM bias among STEM and non-STEM students using the Implicit Relational Assessment Procedure (IRAP). *Journal of Contextual Behavioral Science*, 6(1), 80–90. https://doi.org/10.1016/j.jcbs.2017.02.001

Fiedler, K., Messner, C., & Bluemke, M. (2006). Unresolved problems with the "I", the "A", and the "T": A logical and psychometric critique of the Implicit Association Test (IAT). *European Review of Social Psychology*, 17, 74–147. https://doi.org/10.1080/10463280600681248

Fisher, R. A. (1921). On the 'probable error' of a coefficient of correlation deduced from a small sample. *Metron*, 1, 3–32.

Fisher, R. A., Bennett, J. H. (ed.), & Yates, F. (1990). *Statistical Methods, Experimental Design, and Scientific Inference*. New York, NY: Oxford University Press Inc.

Girod, S., Fassiotto, M., Grewal, D., Ku, M. C., Sriram, N., Nosek, B. A., & Valantine, H. (2016). Reducing implicit gender leadership bias in academic medicine with an educational

intervention. *Academic Medicine*, 91(8), 1143–1150. https://doi.org/10.1097/ACM.0000000000001099

Google Scholar. (2024). Google search of Schweder & Spjøtvoll (1982) paper 'Plots of p-values to evaluate many tests simultaneously'. https://scholar.google.com/scholar?hl=en&as_sdt=0%2C5&q=%22Plots+of+p-values+to+evaluate+many+tests+simultaneously%22&btnG=

Gottfredson, L. S. (2002). Where and why g matters: Not a mystery. *Human performance*, 15(1-2), 25-46. https://doi.org/10.1080/08959285.2002.9668082

Gottfredson, L. S. (2003). *g*, jobs and life. In: Nyborg, H. (ed.), *The Scientific Study of General Intelligence*. New York, NY: Elsevier Sciences Ltd.

Greenwald, A. G., McGhee, D. E., & Schwartz, J. L. K. (1998). Measuring individual differences in implicit cognition: The implicit association test. *Journal of Personality and Social Psychology*, 74, 1464–1480. https://doi.org/10.1037//0022-3514.74.6.1464

Greenwald, A. G., & Farnham, S. D. (2000). Using the implicit association test to measure self-esteem and self-concept. *Journal of Personality and Social Psychology*, 79, 1022–1038. https://doi.org/10.1037//0022-3514.79.6.1022

Haier, R. J. (2009). Cognition and the brain: Sex matters. In: Sommers, C. H. (Ed.), *The Science on Women and Science*. Washington, DC: The AEI Press. 330 pp.

Hanushek, E. A., Schwerdt, G., Wiederhold, S., & Woessmann, L. (2015). "Returns to skills around the world: Evidence from PIAAC." *European Economic Review*, 74, 105−130. https://doi.org/10.1016/j.euroecorev.2014.10.006

Hart, S. A., Petrill, S. A., Thompson, L. A., & Plomin, R. (2009). The ABCs of math: A genetic analysis of mathematics and its links with reading ability and general cognitive ability. *Journal of Educational Psychology*, 101(2), 388–402. https://doi.org/10.1037/a0015115

Hirukawa, M., Murtazashvili, I., & Prokhorov, A. (2023). Yet another look at the omitted variable bias. *Econometric Reviews*, 42(1), 1−27. https://doi.org/10.1080/07474938.2022.2157965

Hui, K., Sukhera, J., Vigod, S., Taylor, V. H., & Zaheer, J. (2020). Recognizing and addressing implicit gender bias in medicine. *Canadian Medical Association Journal*, 192(42), E1269–E1270. https://doi.org/10.1503/cmaj.200286

Hur, M., Ware, R. L., Park, J., McKenna, A. M., Rodgers, R. P., Nikolau, B. J., et al. (2018). Statistically significant differences in composition of petroleum crude oils revealed by volcano plots generated from ultrahigh resolution Fourier transform ion cyclotron resonance mass spectra. Energy & Fuels, 32(2), 1206−1212. https://doi.org/10.1021/acs.energyfuels.7b03061

Hyde, J. S. (2005). The gender similarities hypothesis. *American Psychologist*, 60(6), 581–592. https://doi.org/10.1037/0003-066X.60.6.581

Hyde, J. S. & Mertz, J. E. (2009). Gender, culture and mathematics performance. *Proceedings of the National Academy of Sciences*, 106(22), 8801–8807. https://doi.org/10.1073/pnas.0901265106

Hyde, J. S., Bigler, R. S., Joel, D., Tate, C. C., & van Anders, S. M. (2019). The future of sex and gender in psychology: Five challenges to the gender binary. *American Psychologist*, 74(2), 171–193. https://doi.org/10.1037/amp0000307

Ioannidis, J P. (2005). Why most published research findings are false. *PLoS Medicine*, 2(8), e124. https://doi.org/10.1002/10.1371/journal.pmed.0020124

Ioannidis, J P. (2022). Correction: Why Most Published Research Findings Are False. *PLoS Medicine*, 19(8), e1004085. https://doi.org/10.1371/journal.pmed.1004085. Erratum for: *PLoS Medicine*, 2(8), e124.

Jena, A. B., Khullar, D., Ho, O., Olenski, A. R., & Blumenthal, D. M. (2015). Sex differences in academic rank in US Medical Schools in 2014. *Journal of the American Medical Association*, 314(11), 1149–1158. https://doi.org/10.1001/jama.2015.10680

Kahn, L. M. (2018). Permanent jobs, employment protection, and job content. *Industrial Relations: A Journal of Economy and Society*, 57(3), 469−538. https://doi.org/10.1111/irel.12209

Kindzierski, W., Young, S., Meyer, T., & Dunn, J. D. (2021). Evaluation of a meta-analysis of ambient air quality as a risk factor for asthma exacerbation. *Journal of Respiration*, 1, 173−196. https://doi.org/10.3390/jor1030017

Kuncel, N. R., & Hezlett, S. A. (2010). Fact and fiction in cognitive ability testing for admissions and hiring decisions. *Current Directions in Psychological Science*, 19(6), 339−345. https://doi.org/10.1177/0963721410389459

Kurdi, B., Seitchik, A. E., Axt, J. R., Carroll, T. J., Karapetyan, A., Kaushik, N., et al. (2019). Relationship between the Implicit Association Test and intergroup behavior: A meta-analysis. *American Psychologist*. Advance online publication. http://doi.org/10.1037/amp0000364. Data and other materials are at Open Science Framework (OSF) at https://osf.io/47xw8/.

Kurdi, B., Banaji, M. R. (2019). Relationship between the Implicit Association Test and explicit measures of intergroup cognition: Data from the meta-analysis by Kurdi et al. (2018). psyarxiv.com.

Kyoung Ro, H., Lattuca, L. R., & Alcott, B. (2017). Who goes to graduate school? Engineers' math proficiency, college experience, and self-assessment of skills. *Journal of Engineering Education*, 106(1), 98−122. https://doi.org/10.1002/jee.20154

Levy, J. & Kimura, D. (2009). Women, men, and the sciences. In: Sommers, C. H. (Ed.), *The Science on Women and Science*. Washington, DC: The AEI Press. 330 pp.

Li, W., Freudenberg, J., Suh, Y. J., & Yang, Y. (2014). Using volcano plots and regularized-chi statistics in genetic association studies. Computational Biology and Chemistry, 48, 77−83. https://doi.org/10.1016/j.compbiolchem.2013.02.003

Mitchell, G., & Tetlock, P. E. (2017). Popularity as a poor proxy for utility: The case of implicit prejudice. In: S. O. Lilienfeld & I. D. Waldman (eds.), *Psychological Science Under Scrutiny: Recent Challenges and Proposed Solutions*, (pp. 164–195). Hoboken, NJ: Wiley.

Mitchell, P. G., & Tetlock, P. E. (2020). Stretching the limits of science: Was the implicit-bias debate social psychology's bridge too far? In: J. Krosnick et al. (eds.), *Implicit Bias Theory and Research*. New York, NY: Cambridge University Press.

Morris, M. L. (2016). Vocational interests in the United States: Sex, age, ethnicity, and year effects. *Journal of Counseling Psychology*, 63(5), 604–615. https://doi.org/10.1037/cou0000164

National Science Foundation, National Center for Science and Engineering Statistics (NSF NCSES). (2019). *Women, Minorities, and Persons with Disabilities in Science and Engineering: 2019*. Special report NSF 19-304. https://ncses.nsf.gov/pubs/nsf19304/data

Oswald, F. L., Mitchell, G., Blanton, H., Jaccard, J., & Tetlock, P. E. (2013). Predicting ethnic and racial discrimination: A meta-analysis of IAT criterion studies. *Journal of Personality and Social Psychology*, 105, 171–192. https://doi.org/10.1037/a0032734

Oswald, F. L., Mitchell, G., Blanton, H., Jaccard, J., & Tetlock, P. E. (2015). Using the IAT to predict ethnic and racial discrimination: Small effect sizes of unknown societal significance. *Journal of Personality and Social Psychology*, 108(4), 562–571. https://doi.org/10.1037/pspa0000023

Rounds, J., & Su, R. (2014). The nature and power of interests. *Current Directions in Psychological Science*, 23(2), 98−103. https://doi.org/10.1177/0963721414522812

Schimmack, U. (2019). The implicit association test: A method in search of a construct. *Perspectives on Psychological Science*, 16(2), 396–414. https://doi.org/10.1177/1745691619863798

Schimmack, U. (2021). Invalid claims about the validity of Implicit Association Tests by prisoners of the implicit social-cognition paradigm. *Perspectives on Psychological Science*, 16, 435–442. https://doi.org/10.1177/1745691621991860

Schweder, T., & Spjøtvoll, E. (1982). Plots of p-values to evaluate many tests simultaneously. *Biometrika*, 69, 493−502. https://doi.org/10.1093/biomet/69.3.493

Stewart-Williams, S., & Halsey, L. G. (2018). Men, women, and science: Why the differences and what should be done (preprint). PsyArXiv. https://doi.org/10.31234/osf.io/ms524

Stewart-Williams, S., & Halsey, L. G. (2021). Men, women and STEM: Why the differences and what should be done? *European Journal of Personality*, 35(1), 3–39. https://doi.org/10.1177/0890207020962326

Stoet, G., & Geary, D. C. (2022). Sex differences in adolescents' occupational aspirations: Variations across time and place. *PLoS One*, 17(1), e0261438. https://doi.org/10.1371/journal.pone.0261438

Su, R., Rounds, J., & Armstrong, P., I. (2009). Men and things, women and people: A meta-analysis of sex differences in interests. *Psychological Bulletin*, 135, 859–884. https://doi.org/10.1037/a0017364

Su, R., & Rounds, J. (2015). All STEM fields are not created equal: People and things interests explain gender disparities across STEM fields. *Frontiers in Psychology*, 6, 189. https://doi.org/10.3389/fpsyg.2015.00189

Tan, L., Bradburn, I. S., Knight, D. B., Kinoshita, T., & Grohs, J. (2022). SAT patterns and engineering and computer science college majors: an intersectional, state-level study. *International Journal of STEM Education*, 9(1), 68. https://doi.org/10.1186/s40594-022-00384-6

Tetlock, P. E., & Mitchell, G. (2009). Implicit bias and accountability systems: What must organizations do to prevent discrimination? *Research in Organizational Behavior*, 29, 3–38. https://doi.org/10.1016/j.riob.2009.10.002

Van den Brink, M. (2011). Scouting for talent: Appointment practices of women professors in academic medicine. *Social Science & Medicine*, 72(12), 2033–2040. https://doi.org/10.1016/j.socscimed.2011.04.016

Wicklin, R. (2017). Fisher's transformation of the correlation coefficient. *SAS Blogs*, 20 September 2017. https://blogs.sas.com/content/iml/2017/09/20/fishers-transformation-correlation.html

Wilms, E., Mäthner, E., Winnen, L., & Lanwehr, R. (2021). Omitted variable bias: A threat to estimating causal relationships. *Methods in Psychology*, 5, 100075. https://doi.org/10.1016/j.metip.2021.100075

Young, S. S. (2008). Statistical Analyses and Interpretation of Complex Studies. *Medscape*, Newark, NJ. http://www.medscape.com/viewarticle/571523

Young, S. S., & Kindzierski, W. B. (2019). Evaluation of a meta-analysis of air quality and heart attacks, a case study. *Critical Reviews in Toxicology*, 49(1), 85–94. https://doi.org/10.1080/10408444.2019.1576587

Young, S. S., & Kindzierski, W. B. (2022). Statistical reliability of a diet disease association meta-analysis. *International Journal of Statistics and Probability*, 11(3), 40−50. https://doi.org/10.5539/ijsp.v11n3p40

Young, S. S., & Kindzierski, W. B. (2023). Reproducibility of health claims in meta-analysis studies of COVID quarantine (stay-at-home) orders. *International Journal of Statistics and Probability*, 12(1), 54–65. https://doi.org/10.5539/ijsp.v12n1p54

Young, S. S., & Kindierski, W. B. (2024a). Reproducibility of Implicit Association Test (IAT) – Case study of meta-analysis of racial bias research claims. *International Journal of Statistics and Probability* (accepted, in press).

Young, S. S., & Kindierski, W. B. (2024b). Protocol: Evaluation of gender IAT reliability. Submitted 2024-02-20. https://researchers.one/articles/24.02.00004

Young, S. S., Kindzierski, W., & Randall, D. (2021). *Shifting Sands Unsound Science and Unsafe Regulation, Report #1: Keeping Count of Government Science: P-Value Plotting, P-Hacking, and PM2.5 Regulation*. New York, NY: National Association of Scholars. https://www.nas.org/report-series/shifting-sands-keeping-count-of-government-science

Zitelny, H., Shalom, M., & Bar-Anan, Y. (2017). What is the implicit gender-science stereotype? Exploring correlations between the gender-science IAT and self-report measures. *Social Psychological and Personality Science*, 8(7), 719–735. https://doi.org/10.1177/1948550616683017